\documentclass[conf]{IEEEtran}
\usepackage{graphicx, amsmath, amsfonts, epsfig}
\usepackage{psfrag}
\usepackage{dsfont}
\usepackage{color}
\usepackage{cancel}
\usepackage{subcaption}
\usepackage[font=small]{caption}
\usepackage{multirow}
\usepackage{nicefrac}
\usepackage{manfnt}
\usepackage{amssymb}
\usepackage{epsfig}
\usepackage{exscale}
\usepackage{verbatim}
\usepackage{amstext}
\usepackage{latexsym}
\usepackage{ifthen}
\usepackage{fancybox}
\usepackage{cancel}
\usepackage{cite}
\usepackage{hyperref}
\usepackage{gensymb}

\usepackage{tikz}
\usepackage{pgfplotstable}
\usepackage{rotate}
\usepgfplotslibrary{units}
\usetikzlibrary{%
patterns,%
calc,%
fit,%
arrows,%
plotmarks,%
shadows,%
chains,%
shapes%
}

\tikzstyle{dot} = [draw, circle, minimum size=0.2pt,scale=0.3,fill=black,black]

\usepackage[outline]{contour}
\contourlength{1.2pt}

\newtheorem{theorem}{Theorem}
\newtheorem{lemma}{Lemma}

\newcommand{\I}[0]{\ensuremath{\boldsymbol{I}}}
\newcommand{\X}[0]{\ensuremath{\boldsymbol{X}}}

\newcommand{\R}[0]{\ensuremath{\boldsymbol{R}}}
\newcommand{\Y}[0]{\ensuremath{\boldsymbol{Y}}}
\newcommand{\Z}[0]{\ensuremath{\boldsymbol{Z}}}
\newcommand{\x}[0]{\ensuremath{\boldsymbol{x}}}
\renewcommand{\d}[0]{\ensuremath{\boldsymbol{d}}}
\newcommand{\y}[0]{\ensuremath{\boldsymbol{y}}}
\newcommand{\z}[0]{\ensuremath{\boldsymbol{z}}}
\newcommand{\V}[0]{\ensuremath{\boldsymbol{V}}}

\newcommand{\vu}[0]{\ensuremath{\boldsymbol{u}}}
\renewcommand{\v}[0]{\ensuremath{\boldsymbol{v}}}

\renewcommand{\H}[0]{\ensuremath{\boldsymbol{H}}}
\newcommand{\G}[0]{\ensuremath{\boldsymbol{G}}}
\newcommand{\W}[0]{\ensuremath{\boldsymbol{W}}}
\newcommand{\T}[0]{\ensuremath{\boldsymbol{T}}}

\begin{document}
\IEEEoverridecommandlockouts
\title{On the Degrees-of-Freedom of the MIMO Three-Way Channel with Intermittent Connectivity}
\author{
\IEEEauthorblockN{Anas Chaaban, Aydin Sezgin, and Mohamed-Slim Alouini}
\IEEEauthorblockA{}
\thanks{%
A. Chaaban and M.-S. Alouini are with the Division of Computer, Electrical, and Mathematical Sciences and Engineering (CEMSE) at King Abdullah University of Science and Technology (KAUST), Thuwal, Saudi Arabia. Email: \{anas.chaaban,slim.alouini\}@kaust.edu.sa.

A. Sezgin is with the Institute of Digital Communication Systems at the Ruhr-Universit\"at Bochum, Bochum, Germany. Email: aydin.sezgin@rub.de.
}
}

\maketitle

\begin{abstract}
The degrees-of-freedom (DoF) of the multi-antenna three-way channel (3WC) with an intermittent node is studied. Special attention is given to the impact of adaptation. A nonadaptive transmission scheme based on interference alignment, zero-forcing, and erasure-channel treatment is proposed, and its corresponding DoF region is derived. Then, it is shown that this scheme achieves the sum-DoF of the intermittent channel, in addition to the DoF region of the nonintermittent one. Thus, adaptation is not necessary from those perspectives. To the contrary, it is shown that adaptation is necessary for achieving the DoF region of the intermittent case. This is shown by deriving an outer bound for the intermittent channel with nonadaptive encoding, and giving a counterexample of an adaptive scheme which achieves DoF tuples outside this bound. This highlights the importance of cooperation in this intermittent network.
\end{abstract}

\section{Introduction}
Multi-way, full-duplex, and device-to-device (D2D) communications are important techniques that are expected to gain more prominence in future communication systems. Multi-way communication refers to communication between multiple nodes each acting as a source, a destination, and possibly a relay. Full-duplex operation is defined as when these three functionalities take place over the same time/frequency resources, and D2D communication refers to direct communication between users without or with limited base-station intervention. Those techniques attracted and continue to attract research attention \cite{ChaabanSezgin_FnT,SongDevroyeShaoNgo,ChengDevroyeLiu,SabharwalSchniterGuo,TehraniUysalYanikomeroglu}.

Consider a setup where three nodes (D2D users e.g.) communicate with each other in a multi-way fashion. This setup can be modeled as a three-way channel (3WC), an extension of Shannon's two-way channel \cite{Shannon_TWC} which has been studied in \cite{ChaabanMaierSezginMathar,Ong,ElmahdyKeyiMohassebElBatt}. Therein, it is assumed that the nodes are always connected. This assumption is not always valid in practice since a node might have intermittent connectivity, e.g. due to shadowing, or if a D2D node causes strong interference to a cellular user, in which case it is not permitted to use its band~\cite{mach2015band}.

The impact of intermittency on the performance of various networks was studied in \cite{KarakusWangDiggavi,WangSuhDiggaviViswanath,VahidMaddahAliAvestimehr,YehWang} for instance. In this paper, we study the impact of intermittency on the multiple-input multiple-output (MIMO) 3WC. We consider a full-duplex MIMO 3WC with full message-exchange, where each node has an independent message to each of the other two nodes. The permanent nodes have only causal knowledge of the availability of the intermittent node, which can be obtained by estimating its activity from the received signals. For this model, we study the degrees-of-freedom (DoF), i.e., the capacity scaling versus signal-to-noise ratio (SNR) in a dB scale. We pay particular attention to the necessity or the lack thereof, of {\it adaptive encoding} where the transmit signal of each node is allowed to depend on its previously received signals. This issue has been studied for various channels earlier~\cite{Varshney,ChengDevroye,SongAlajajiLinder}.

First, we devise a nonadaptive scheme based on interference alignment and zero-forcing, where intermittency is treated as an erasure channel, and we derive its achievable DoF region. Then, we derive DoF upper bounds that prove that this scheme achieves the sum-DoF of the channel. It follows that {\it as far as the sum-DoF is concerned, adaptation is not necessary}. 
This scheme also achieves the DoF region of the channel without intermittency, and hence, {\it for the nonintermittent channel, adaptation is not necessary for achieving the DoF region}. After showing the unnecessity of adaptation in those two cases, we prove that {\it adaptation is necessary to achieve the DoF region of the intermittent channel}. To show this, we derive a DoF outer bound that holds under nonadaptive encoding, and provide an adaptive scheme that achieves rates that violate this outer bound. This proves that adaptation enlarges the DoF region in the intermittent 3WC.

Throughout the paper, we use $x_i^n$ for some $i$ to denote $(x_{i,1},\ldots,x_{i,n})$. The $N\times N$ identity matrix is denoted $\I_N$. We write $\X\sim\mathcal{CN}(\boldsymbol{0},\boldsymbol{Q})$ to indicate that $\X$ is a complex Gaussian random variable with zero mean and covariance matrix $\boldsymbol{Q}$. We write $x^+$ to denote $\max\{0,x\}$ for some $x\in\mathbb{R}$, $\|\x\|_i$ to denote the $\ell_i$-norm of $\x$, and $\H^\dagger$, $\H^H$, and ${\rm span}(\H)$ to denote the pseudo-inverse, the Hermitian transpose, and the the subspace spanned by the columns of~$\H$.

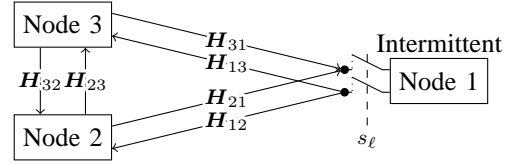
\begin{figure}[t]
\centering
\begin{tikzpicture}
\node (n2) at (0,0) [rectangle,draw,minimum height=.6cm] {Node 2};
\node (n3) at (0,1.5) [rectangle,draw,minimum height=.6cm] {Node 3};
\node (n1) at (5,.75) [rectangle,draw,minimum height=.6cm] {Node 1};

\draw[-] ($(n1.west)+(0,.15)$) to ($(n1.west)+(-.1,.15)$) to ($(n1.west)+(-.5,.35)$);
\draw[dotted]  ($(n1.west)+(-.5,.35)$) to ($(n1.west)+(-.5,.15)$);
\node (p1) at ($(n1.west)+(-.6,.15)$) [dot] {};
\draw[-]  ($(n1.west)+(-.5,.15)$) to (p1);
\draw[-] ($(n1.west)+(0,-.15)$) to ($(n1.west)+(-.1,-.15)$) to ($(n1.west)+(-.5,.05)$);
\draw[dotted]  ($(n1.west)+(-.5,.05)$) to ($(n1.west)+(-.5,-.15)$);
\node (p2) at ($(n1.west)+(-.6,-.15)$) [dot] {};
\draw[-]  ($(n1.west)+(-.5,-.15)$) to (p2);

\draw[->] ($(n2.east)+(0,.15)$) to node {\footnotesize \contour{white}{$\H_{21}$}} (p1);
\draw[<-] ($(n2.east)-(0,.15)$) to node {\footnotesize \contour{white}{$\H_{12}$}} (p2);
\draw[->] ($(n3.east)+(0,.15)$) to node {\footnotesize \contour{white}{$\H_{31}$}} (p1);
\draw[<-] ($(n3.east)-(0,.15)$) to node {\footnotesize \contour{white}{$\H_{13}$}} (p2);
\draw[->] ($(n2.north)+(0.3,0)$) to node {\footnotesize \contour{white}{$\H_{23}$}} ($(n3.south)+(.3,0)$);
\draw[<-] ($(n2.north)-(0.3,0)$) to node {\footnotesize \contour{white}{$\H_{32}$}} ($(n3.south)-(.3,0)$);

\node at ($(n1)+(0,.5)$) {Intermittent};
\node (sl) at ($(n1.west)+(-.3,-.8)$) {\footnotesize $s_\ell$};
\draw[dashed] (sl) to ($(sl)+(0,1.2cm)$);

\end{tikzpicture}
\caption{A MIMO 3WC with an intermittent node: $\H_{ij}$ is the channel matrix and $s_\ell\in\{0,1\}$ is the intermittency-state at time instant~$\ell$.}
\label{Fig:3WC}
\end{figure}

\section{System Model}
\label{Model}
Consider a system where three MIMO full-duplex nodes communicate in a multi-way manner using the same medium, with one of the nodes being intermittently available (Fig. \ref{Fig:3WC}). For some transmission duration $n\in\mathbb{N}$ (in symbols), let $s^n$ denote the intermittency of node 1, where for $\ell\in\{1,\ldots,n\}$, $s_\ell=1$ means that node 1 is available, and $s_\ell=0$ otherwise. The state $s^n$ is a sequence of independent and identically distributed (i.i.d) Bernoulli random variables $S_\ell$ with probability $\mathbb{P}(s_\ell=1)=\tau$ and $\mathbb{P}(s_\ell=0)=1-\tau\triangleq\bar{\tau}$. This sequence is known at node 1. However, knowledge of $s_\ell$ is only available causally at node $i\in\{2,3\}$, i.e., node $i$ does not know $s_\ell$ at the beginning of the $\ell$-th transmission, and can only obtain it after receiving the $\ell$-th received signal from which the activity of node 1 can be detected with certainty.

Node $i\in\{1,2,3\}$ is equipped with $M_i$ transmit and receive antennas. Its transmit signal at time index $\ell$ is represented by $\x_{i,\ell}$; a realization of a random vector $\X_{i,\ell}\in\mathbb{C}^{M_i\times 1}$ that satisfies a power constraint\footnote{Any power discrepancy is absorbed into the channel gains.} 
$\sum_{\ell=1}^n\mathbb{E}[\|\X_{i,\ell}\|_2^2]\leq nP$. 
Clearly $\x_{1,\ell}=\boldsymbol{0}$ if $s_\ell=0$. The received signals are 
\begin{align}
\y_{1,\ell}&=\H_{21}\x_{2,\ell}+\H_{31}\x_{3,\ell}+\z_{1,\ell},\quad  \text{if }s_\ell=1\\
\y_{2,\ell}&=\H_{12}\x_{1,\ell}+\H_{32}\x_{3,\ell}+\z_{2,\ell},\\
\y_{3,\ell}&=\H_{13}\x_{1,\ell}+\H_{23}\x_{2,\ell}+\z_{3,\ell},
\end{align}
and $\y_{1,\ell}=\boldsymbol{0}$ if $s_\ell=0$, where $\H_{ji}\in\mathbb{C}^{M_j\times M_i}$ and $\H_{ki}\in\mathbb{C}^{M_k\times M_i}$ represent the channel matrices from nodes $j$ and $k$ to node $i$, respectively, and $\z_{i,\ell}$ is a realization of $\Z_{i,\ell}\sim\mathcal{CN}(\boldsymbol{0},\sigma^2\I_{M_i})$, i.i.d. with respect to $\ell$. 

We denote $P/\sigma^2$ by $\rho$ and call it SNR throughout the paper. We assume without loss of generality that $M_2\geq M_3$. We also assume that $M_1\geq M_2$. The channel matrices are generated randomly from a continuous distribution, held constant throughout the transmission, and are known globally. The message sets, encoding, and decoding, and achievability are defined in the standard Shannon sense \cite{CoverThomas}. The encoder at node $i$, $\mathcal{E}_{i,\ell}$, can be either adaptive where dependence of $\x_{i,\ell}$ on $\y_i^{\ell-1}$ is allowed, or restricted (nonadaptive) where it is not. 
These possibilities are shown in Table \ref{Tab:Enc}. 

\begin{table}[t]
\centering
\begin{tabular}{c||c|c}
 & $\x_{1,\ell}$ & $\x_{i,\ell}$,\ $i\in\{2,3\}$\\\hline
Restricted & $\mathcal{E}_{1,\ell}(w_{12},w_{13},s^n)$ & $\mathcal{E}_{i,\ell}(w_{ij},w_{ik})$\\\hline
Adaptive & $\mathcal{E}_{1,\ell}(w_{12},w_{13},s^n,\y_{1}^{\ell-1})$ & $\mathcal{E}_{i,\ell}(w_{ij},w_{ik},s^{\ell-1},\y_{i}^{\ell-1})$
\end{tabular}
\caption{The encoding possibilities. Here $w_{ij}$ is the message to be sent from node $i$ to node $j$.}
\label{Tab:Enc}
\end{table}



The DoF region is the set of achievable DoF tuples $\d=(d_{12},d_{13},d_{21},d_{23},d_{31},d_{32})\in\mathbb{R}_+^6$ defined as in \cite{JafarShamai_XChannel}. Roughly speaking, if a rate tuple (function of $\rho$)
\begin{align*}
\R(\rho)&=\left(R_{12}(\rho),R_{13}(\rho),R_{21}(\rho),R_{23}(\rho),R_{31}(\rho),R_{32}(\rho)\right)
\end{align*}
where $R_{ij}(\rho)$ is the rate of the message from node $i$ to node $j$, is achievable, then the DoF tuple $\d$ with $d_{ij}=\lim\sup_{\rho\to\infty}\frac{R_{ij}(\rho)}{\log(\rho)}$ is achievable. We denote the DoF region under restricted encoding and adaptive encoding for a given $\tau$ by $\mathcal{D}_{{\rm r},\tau}$ and $\mathcal{D}_{{\rm a},\tau}$, respectively, and we define the sum-DoF as $d_{{\rm r},\tau}=\max_{\d\in\mathcal{D}_{{\rm r},\tau}} \|\d\|_1$ and $d_{{\rm a},\tau}=\max_{\d\in\mathcal{D}_{{\rm a},\tau}} \|\d\|_1$.

Next, we describe a restricted transmission scheme, and we derive its achievable DoF region.

\section{Restricted Encoding Transmission Scheme}
\label{Sec:Restricted}
In this section, we prove the following theorem.

\begin{theorem}
\label{Thm:Restricted}
The DoF region of the 3WC satisfies $\mathcal{D}_{\rm r,\tau}^{[\rm in]}\subseteq\mathcal{D}_{\rm r,\tau}\subseteq\mathcal{D}_{\rm a,\tau}$, where the achievable inner bound $\mathcal{D}_{\rm r,\tau}^{[\rm in]}$ is the set of $\d\in\mathbb{R}_+^6$ satisfying the following for $i,j\in\{2,3\}$, $i\neq j$:
\begin{align}
d_{1i}+d_{1j}+\tau d_{ij}&\leq \tau M_1,&d_{31}+\tau d_{32}&\leq \tau M_3,\\
d_{i1}+d_{j1}+\tau d_{ij}&\leq \tau M_1,&d_{13}+\tau d_{23}&\leq \tau M_3,\\
d_{i1}+d_{1j}+\tau d_{ij}&\leq \tau M_2.
\end{align}
\end{theorem}
\begin{IEEEproof}
The inclusion of $\mathcal{D}_{\rm r,\tau}$ in $\mathcal{D}_{\rm a,\tau}$ is obvious. The achievability of $\mathcal{D}_{\rm r,\tau}^{[\rm in]}$ is proved in the rest of this section.
\end{IEEEproof}

Note that the factor $\tau$ in the inequalities above imposes a larger penalty on the streams going through the intermittent links. If we interpret $\tau M_3$ in $d_{31}+\tau d_{32}\leq\tau M_3$ as available resources, then increasing $d_{32}$ by $1$ `eats' $\tau$ units of resources, while increasing $d_{31}$ by $1$ `eats' $1$ unit of resources. Thus, transmission between nodes 2 and 3 is `cheaper' by a factor of $\tau$, no matter how large $M_1$ is as we shall see later. Next, we prove the achievability of $\mathcal{D}_{\rm r,\tau}^{[\rm in]}$.


\subsubsection{Encoding}
Each node splits its message $w_{ij}$ into $w_{ij}^{[1]}$ and $w_{ij}^{[2]}$ to be sent using zero-forcing and interference alignment, respectively. Encoding proceeds as follows.

Since node 1 is available for a fraction of time, say $m=\|s^n\|_0$ out of the $n$ transmissions, it encodes $w_{12}^{[q]}$ and $w_{13}^{[q]}$, $q\in\{1,2\}$, into codewords $\vu_{12}^{[q]m}$ and $\vu_{13}^{[q]m}$ with i.i.d. $\mathcal{CN}(\boldsymbol{0},p_1\I_{a_{12}^{[q]}})$ and $\mathcal{CN}(\boldsymbol{0},p_1\I_{a_{13}^{[q]}})$ symbols, respectively. Here $a_{ij}^{[q]}$ is the vector length, and $p_1$ is the power of each component of $\vu_{12}^{[q]m}$ and $\vu_{13}^{[q]m}$. Then, those codewords are extended to length $n$ codewords $\x_{12}^{[q]n}$ and $\x_{13}^{[q]n}$ by inserting zeros where $s_\ell=0$ (note that $s^n$ is known at node 1).

Now, nodes 2 and 3 are available all the time, but they do not have apriori knowledge of $s_\ell$. Thus, these nodes use standard random Gaussian codebooks to encode their messages, and treat the channel to node 1 as an erasure channel with erasure probability $\bar{\tau}$. Node 2 encodes $w_{21}^{[q]}$ and $w_{23}^{[q]}$, $q\in\{1,2\}$, into codewords $\x_{21}^{[q]n}$ and $\x_{23}^{[q]n}$ with i.i.d. $\mathcal{CN}(\boldsymbol{0},p_2\I_{a_{21}^{[q]}})$ and $\mathcal{CN}(\boldsymbol{0},p_2\I_{a_{23}^{[q]}})$ symbols, respectively. Similarly, node 3 encodes $w_{31}^{[q]}$ and $w_{32}^{[q]}$, $q\in\{1,2\}$, into codewords $\x_{31}^{[q]n}$ and $\x_{32}^{[q]n}$ with i.i.d. $\mathcal{CN}(\boldsymbol{0},p_3\I_{a_{31}^{[q]}})$ and $\mathcal{CN}(\boldsymbol{0},p_3\I_{a_{32}^{[q]}})$ symbols, respectively.

To satisfy the power constraint, the powers are chosen as
\begin{align}
p_1&=(a_{12}^{[1]}+a_{12}^{[2]}+a_{13}^{[1]}+a_{13}^{[2]})^{-1}m^{-1}nP,\\ 
\label{P2P3}
p_i&=(a_{i1}^{[1]}+a_{i1}^{[2]}+a_{ij}^{[1]}+a_{ij}^{[2]})^{-1}P,
\end{align} 
$i,j\in\{2,3\}$, $i\neq j$. This encoding is restricted since it uses neither $s^{\ell-1}$ at nodes 2 and 3, nor $\y_i^{\ell-1}$ at nodes 1, 2, and 3.

\subsubsection{Transmission}
At time $\ell$, node $i$ sends
\begin{align}
\x_{i,\ell}&=\sum_{q=1}^2\left[\V_{ij}^{[q]}\x_{ij,\ell}^{[q]}+\V_{ik}^{[q]}\x_{ik,\ell}^{[q]}\right],
\end{align} 
where $j,k\in\{1,2,3\}\setminus\{i\}$, $j\neq k$, and $\V_{ij}^{[q]}\in\mathbb{C}^{M_i\times a_{ij}^{[q]}}$ is a beamforming matrix. Zero-forcing is achieved by choosing the matrices $\V_{ij}^{[1]}$ so that $\H_{ik}\V_{ij}^{[1]}=\boldsymbol{0}$ for distinct $i,j,k\in\{1,2,3\}$. These matrices exist if
\begin{align}
(M_i-M_k)^+\geq a_{ij}^{[1]},
\end{align}
ensuring that node $i$ has enough antennas to send $a_{ij}^{[1]}$ streams to node $j$ without interfering with node $k$. To avoid any overlap of the transmit signals in the transmit signal space, we require
\begin{align}
\sum_{q=1}^2 (a_{ij}^{[q]}+a_{ik}^{[q]})\leq M_i.
\end{align}

\subsubsection{Decoding}
Node 1 receives $\y_{1,\ell}=\boldsymbol{0}$ if $s_\ell=0$ and 
\begin{align}
\label{y1_1}
\y_{1,\ell}&=\sum_{j=2}^3\sum_{q=1}^2\H_{j1}\V_{j1}^{[q]}\x_{j1,\ell}^{[q]}+\G_{23}\left[\begin{smallmatrix}\x_{23\ell}^{[2]}\\ \x_{32,\ell}^{[2]}\end{smallmatrix}\right]+\z_{1,\ell}
\end{align}
otherwise, where $\G_{23}=[\H_{21}\V_{23}^{[2]},\ \H_{31}\V_{32}^{[2]}]$. This signal consists of four desired signals plus interference. To decode a desired signals, say $\x_{21}^{[1]n}$, node 1 zero-forces the remaining signals by multiplying $\y_1^n$ by a post-coder $\T_{21}^{[1]}\in\mathbb{C}^{a_{21}^{[1]}\times M_1}$ satisfying $\T_{21}^{[1]}\T_{21}^{[1]H}=\I_{a_{21}^{[1]}}$ and
\begin{align}
\label{ZFRx}
\T_{21}^{[1]}[\H_{21}\V_{21}^{[2]},~\H_{31}\V_{31}^{[1]},~\H_{31}\V_{31}^{[2]},~\G_{23}]&=\boldsymbol{0},\\
\label{ZFRankConst}
\mathrm{rank}(\T_{21}^{[1]}\H_{21}\V_{21}^{[1]})&=a_{21}^{[1]}.
\end{align} 
After post-coding, node 1 is left with the signal 
$\y_{21,\ell}=\T_{21}^{[1]}\H_{21}\V_{21}^{[1]}\x_{21,\ell}^{[1]}+\T_{21}^{[1]}\z_{1,\ell}$
if $s_\ell=1$ and $\y_{21,\ell}=\boldsymbol{0}$ otherwise. This is an erasure channel over which the rate $I(\x_{21,\ell}^{[1]};\y_{21,\ell},s_\ell)= I(\x_{21,\ell}^{[1]};\y_{21,\ell}|s_\ell)=\tau I(\x_{21,\ell}^{[1]};\y_{21,\ell}|s_\ell=1)$ is achievable from node 2 to node 1 for $n$ large.\footnote{Recall that $s^n$ is known at the decoding stage.} Since we used Gaussian i.i.d. codes, this rate is
\begin{align}
\tau \log\left|\I_{a_{21}^{[1]}}+\frac{p_2}{\sigma^2}\T_{21}^{[1]}\H_{21}\V_{21}^{[1]}\V_{21}^{[1]H}\H_{21}^H\T_{21}^{[1]H}\right|,
\end{align}
leading to a DoF of $\tau a_{21}^{[1]}$ as long as \eqref{ZFRankConst} is satisfied (cf. \eqref{P2P3}).

A similar procedure can be applied for decoding $\x_{21}^{[2]n}$, $\x_{31}^{[1]n}$, and $\x_{31}^{[2]n}$, to achieve DoF of $\tau a_{21}^{[2]}$, $\tau a_{31}^{[1]}$, $\tau a_{31}^{[2]}$. The existence of the post-coders $\T_{i1}^{[q]}$ which allow this procedure is guaranteed as long as the columns of 
\begin{align}
[\H_{21}\V_{21}^{[1]},~\H_{21}\V_{21}^{[2]},~\H_{31}\V_{31}^{[1]},~\H_{31}\V_{31}^{[2]},~\G_{23}]
\end{align}
are linearly independent. Let $\bar{a}_{23}^{[2]}$ be the dimension of ${\rm span}(\H_{21}\V_{23}^{[2]})\cap{\rm span}(\H_{31}\V_{32}^{[2]})$. Then, ${\rm span}(\G_{23})$ has $a_{23}^{[2]}+a_{32}^{[2]}-\bar{a}_{23}^{[2]}$ dimensions, and the above linear independence is possible if we choose
\begin{align}
\label{RXspace}
\sum_{q=1}^2(a_{21}^{[q]}+a_{31}^{[q]})+a_{23}^{[2]}+a_{32}^{[2]}-\bar{a}_{23}^{[2]}\leq M_1.
\end{align}
To minimize the impact of interference, we choose $\V_{ij}^{[q]}$ so that $\bar{a}_{23}^{[2]}$ is maximized. This can not be chosen arbitrarily large, as it has to be smaller than each of $a_{23}^{[2]}$ and $a_{32}^{[2]}$, and also smaller than the dimension of ${\rm span}(\H_{21})\cap{\rm span}(\H_{31})$, which is $(M_2+M_3-M_1)^+$ almost surely. Thus,
\begin{align}
\label{IntSpace}
\min\{a_{23}^{[2]},a_{32}^{[2]},(M_2+M_3-M_1)^+\}\geq \bar{a}_{23}^{[2]}.
\end{align}

The same arguments can be applied at nodes 2 and 3, for decoding their desired signals. This achieves $\tau a_{12}^{[1]}$, $\tau a_{12}^{[2]}$, $a_{32}^{[1]}$, and $a_{32}^{[2]}$ DoF at node 2, and $\tau a_{13}^{[1]}$, $\tau a_{13}^{[2]}$, $a_{23}^{[1]}$, and $a_{23}^{[2]}$ DoF at node 3, leading to similar constraints as \eqref{RXspace} and \eqref{IntSpace}.

\subsubsection{Achievable DoF Region}
The constraints can be combined as follows
\begin{align}
\label{ZFConstraint}
(M_i-M_k)^+&\geq a_{ij}^{[1]},\\
\label{TXConstraint}
\sum_{q=1}^2 (a_{ij}^{[q]}+a_{ik}^{[q]})&\leq M_i,\\
\label{AConstraint}
\min\{a_{ij}^{[2]},a_{ji}^{[2]},(M_i+M_j-M_k)^+\}&\geq \bar{a}_{ij}^{[2]},\\
\label{RXConstraint}
\sum_{q=1}^2(a_{ji}^{[q]}+a_{ki}^{[q]})+a_{jk}^{[2]}+a_{kj}^{[q]}-\bar{a}_{jk}^{[2]}&\leq M_i.
\end{align}
for distinct $i,j,k\in\{1,2,3\}$, where $\bar{a}_{12}^{[2]}$ and $\bar{a}_{13}^{[2]}$ are the dimensions of ${\rm span}(\H_{13}\V_{12}^{[2]})\cap{\rm span}(\H_{23}\V_{21}^{[2]})$ and ${\rm span}(\H_{12}\V_{13}^{[2]})\cap{\rm span}(\H_{32}\V_{31}^{[2]})$, respectively. By adding the achievable DoF per stream, we obtain $d_{ij}$ (e.g. $d_{21}=\tau a_{21}^{[1]}+\tau a_{21}^{[2]}$). Substituting $d_{ij}$ in \eqref{ZFConstraint}--\eqref{RXConstraint}, using $M_1\geq M_2\geq M_3$ and Fourier Motzkin's elimination leads to the DoF region in Theorem \ref{Thm:Restricted}. Details are omitted due to space limitations.

Next, we study the optimality of this scheme.

\section{Optimality Discussion}
\label{Sec:Optimality}

\subsection{Sum-DoF}
We first consider the sum-DoF of the channel, and start by presenting the following DoF upper bounds.

\begin{lemma}
\label{Lemma:DoF_Bounds}
The following must be satisfied by any DoF tuple $\d\in\mathcal{D}_{\rm a,\tau}$ (and hence also $\d\in\mathcal{D}_{\rm r,\tau}$):
\begin{align}
d_{13}+d_{23}+d_{21}&\leq \tau M_2+\bar{\tau}M_3,\\
d_{31}+d_{32}+d_{12}&\leq \tau M_2+\bar{\tau}M_3.
\end{align}
\end{lemma}
\begin{IEEEproof}
For brevity, we denote $(W_{ij},W_{ik})$ by $\W_i$, and use $\epsilon_{1n}$, $\epsilon_{2n}$, and $\epsilon_{3n}$ to denote quantities that vanish as $n\to\infty$. Let $\tilde{\H}_{23}$ be an $(M_2-M_3)\times M_2$ matrix so that $\hat{\H}_{23}\triangleq[\H_{23}^T,\ \tilde{\H}_{23}^T]^T$ has full rank $M_2$. Such a matrix exists almost surely. Also, let $\tilde{\Y}_{3,\ell}$ be defined as $\tilde{\H}_{23}\X_{2,\ell}+\tilde{\Z}_{3,\ell}$ if $S_\ell=1$ and $\boldsymbol{0}$ otherwise, where $\tilde{\Z}_{3,\ell}\sim\mathcal{CN}(\boldsymbol{0},\sigma_3^2\I_{M_2-M_3})$, and define $\hat{\Y}_{3,\ell}=[\Y_{3,\ell}^T,\ \tilde{\Y}_{3,\ell}^T]^T$. Now, consider any code for the 3WC, and let us establish a bound on $R_{13}+R_{23}+R_{21}$.\footnote{We write $R_{ij}(\rho)$ simply as $R_{ij}$ for brevity.} We give $(\tilde{\Y}_3^n,W_{12})$ and $(\hat{\Y}_3^n,\W_3,W_{23})$ as side information to nodes 3 and 1, respectively. By Fano's inequality, we have
\begin{align*}
n(R_{13}+R_{23}-\epsilon_{1n})&\leq I(W_{13},W_{23};\hat{\Y}_3^n,S^n,\W_3,W_{12}),\\
n(R_{21}-\epsilon_{2n})&\leq I(W_{21};\Y_1^n,\hat{\Y}_3^n,S^n,\W_1,\W_3,W_{23}).
\end{align*}
Recall that each node can estimate $S^n$ with certainty from the received signals as assumed in the system model. Using the chain rule, the independence of the messages of each other and of $S^n$, and combining the two bounds yields
\begin{align*}
n(R_{13}+R_{23}+R_{21}-\epsilon_{3n})&\leq I(\W_2,W_{13};\hat{\Y}_3^n|S^n,\W_3,W_{12})\nonumber\\
&\hspace{-1cm} +I(W_{21};\Y_1^n|\hat{\Y}_3^n,S^n,\W_1,\W_3,W_{23}).
\end{align*}
The second term in this bound is equal to $\sum_{\ell=1}^n I(W_{21};\Y_{1,\ell}|\Y_1^{\ell-1},\hat{\Y}_3^n,S^n,\W_1,\W_3,W_{23})$, which is $no(\log(\rho))$,\footnote{$\lim_{\rho\to\infty}\frac{o(\log(\rho))}{\log(\rho)}=0.$} since given $\W_1$, $\W_3$, $\Y_1^{\ell-1}$, $\Y_3^n$, and $S^n$, and using the adaptive encoder, we can construct a noisy version of $\Y_{1,\ell}$ for all $\ell$ with $S_\ell=1$ given by $\H_{21}\hat{\H}_{23}^{-1}\left[\begin{smallmatrix}\Y_{3,\ell}-\H_{13}\X_{1,\ell}\\\tilde{\Y}_{3,\ell}\end{smallmatrix}\right]+\H_{31}\X_{3,\ell}$. 
On the other hand, using standard steps
\begin{align}
&I(\W_2,W_{13};\hat{\Y}_3^n|S^n,\W_3,W_{12})\nonumber\\
\label{DoF_Sum}
&\leq \sum_{\ell=1}^n I(\X_{1,\ell},\X_{2,\ell};\hat{\Y}_{3,\ell}|S_\ell)\\
&= \sum_{\ell=1}^n \tau I(\X_{1,\ell},\X_{2,\ell};\hat{\Y}_{3,\ell}|S_\ell=1)+\bar{\tau} I(\X_{2,\ell};\Y_{3,\ell}|S_\ell=0)\nonumber\\
&\leq n(\tau M_2+\bar{\tau}M_3)\log(\rho)+no(\log(\rho)),
\end{align}
since the first and second terms represent $(M_1+M_2)\times M_2$ and $M_2\times M_3$ MIMO channels with $M_2$ and $M_3$ DoF almost surely ($M_3\leq M_2$), respectively. Combining terms, dividing by $n$ and letting $n\to\infty$, this yields the bound
\begin{align*}
R_{13}+R_{23}+R_{21}\leq (\tau M_2+\bar{\tau}M_3)\log(\rho)+o(\log(\rho)),
\end{align*}
which consequently leads to the first DoF bound. The second is obtained similarly by giving $(\Y_2^n,\W_2,W_{32})$ and $W_{13}$ as side information to nodes 1 and 2, respectively.
\end{IEEEproof}

Based on Lemma \ref{Lemma:DoF_Bounds}, we can state the following theorem.

\begin{theorem}
The sum-DoF of the intermittent 3WC is given by $d_{\rm r,\tau}=d_{\rm a,\tau}=2\tau M_2+2\bar{\tau}M_3$.
\end{theorem}
\begin{IEEEproof}
Achievability follows from Theorem \ref{Thm:Restricted} by using the simplex method \cite{MatousekGartner} to maximize the sum-DoF subject to the DoF constraints. In particular, it follows by setting $a_{12}^{[1]}=a_{21}^{[1]}=M_2-M_3$, $a_{23}^{[2]}=a_{32}^{[2]}=M_3$, and $a_{12}^{[2]}=a_{21}^{[2]}=a_{23}^{[1]}=a_{32}^{[1]}=d_{13}=d_{31}=0$ in the scheme described in Sec. \ref{Sec:Restricted}. The converse follows by adding the DoF bounds in Lemma~\ref{Lemma:DoF_Bounds}.
\end{IEEEproof}

This agrees with intuition. To maximize the sum-DoF, one should capitalize on the stable links between nodes 2 and 3, and use any remaining resources for communicating with the intermittent node 1. This theorem proves that {\it adaptation is not necessary for achieving the sum-DoF of the intermittent 3WC}. The same does not hold true from a DoF region perspective as we shall see next.

\subsection{DoF Region}
In this section, we show that adaptation is necessary for achieving the DoF region of the intermittent 3WC. This result is particularly interesting in light of the following statement.

\begin{theorem}
The scheme in Sec. \ref{Sec:Restricted} achieves the DoF region of the nonintermittent 3WC ($\tau=1$) given by $\mathcal{D}_{\rm a,1}=\mathcal{D}_{\rm r,1}^{[\rm in]}$.
\end{theorem}
\begin{IEEEproof}
The proof is based on upper bounds in \cite{MaierChaabanMathar,ElmahdyKeyiMohassebElBatt}, and is omitted for lack of space.
\end{IEEEproof}

Therefore, from a DoF-region point-of-view, {\it adaptation is not necessary in the nonintermittent case}.\footnote{Adaptation is still necessary from an achievable rate point-of-view \cite{ChaabanMaierSezginMathar}, but the gain of adaptation does not scale with $\rho$.} Interestingly, the same is not true in the intermittent case. To prove this, first we need a DoF outer bound for the {\it restricted} intermittent 3WC, and second, we need an adaptive scheme which achieves DoF tuples outside this outer bound. The first step is tackled in the following lemma. 

\begin{lemma}
\label{Lemma:Out}
Under restricted encoding, we have $\mathcal{D}_{\rm r,\tau}\subset \mathcal{D}_{\rm r,\tau}^{[\rm out]}$ defined as the set of $\d\in\mathbb{R}_+^6$ satisfying $d_{31}+\tau d_{32}\leq \tau M_3$.
\end{lemma}
\begin{IEEEproof}
Let $\breve{\Y}_{2,\ell}=S_\ell\Y_{2,\ell}$ and let us give $(\breve{\Y}_2^n,\W_2)$ to node 1 as side information. From Fano's inequality, we have 
\begin{align}
n(R_{31}-\epsilon_{1n})
&\leq I(W_{31};\Y_1^n,\breve{\Y}_2^n|S^n,\W_1,\W_2)\\
&= I(W_{31};\breve{\Y}_2^n|S^n,\W_1,\W_2)\nonumber\\
&\quad + I(W_{31};\Y_1^n|S^n,\W_1,\W_2,\breve{\Y}_2^n).
\end{align}
Given $\breve{\Y}_2^n$, $\W_1$, $\W_2$, and $S^n$, we can construct a noisy version of $\Y_1^n$ for $S_\ell=1$ given by $\H_{31}\H_{32}^\dagger(\breve{\Y}_{2,\ell}-\H_{12}\X_{1,\ell})+\H_{21}\X_{2,\ell}$, where $\H_{32}^\dagger$ exists almost surely. Thus, $I(W_{31};\Y_1^n|S^n,\W_1,\W_2,\breve{\Y}_2^n)=n\tau o(\log(\rho))$, and hence
\begin{align}
&n(R_{31}-\epsilon_{1n}-\tau o(\log(\rho)))
\leq I(W_{31};\breve{\Y}_2^n|S^n,\W_1,\W_2)\nonumber\\
&\quad = \tau\sum_{\ell=1}^n I(W_{31};\Y_{2,\ell}|S^n,\W_1,\W_2,\breve{\Y}_2^{\ell-1},S_\ell=1).
\end{align}
On the other hand, giving $(\W_1,W_{31})$ to node 3 as side information and using Fano's inequality, we have
\begin{align}
n(\tau R_{32}-\epsilon_{2n})
&\leq \tau I(W_{32};\Y_2^n|S^n,\W_1,\W_2,W_{31})\\
&= \tau \sum_{\ell=1}^n [h(\Y_{2,\ell}|S^n,\W_1,\W_2,W_{31},\Y_2^{\ell-1})\nonumber\\
&\quad -h(\Y_{2,\ell}|S^n,\W_1,\W_2,\W_3,\Y_2^{\ell-1})].\nonumber
\end{align}
Since conditioning does not increase entropy, the first term is upper bounded by $h(\Y_{2,\ell}|S^n,\W_1,\W_2,W_{31},\breve{\Y}_2^{\ell-1})$. Moreover, since restricted encoding can be used to generate $\X_1^n$ and $\X_3^n$ from $\W_1$ and $\W_3$, the second entropy term is equal to $h(\Z_{2,\ell})=h(\Y_{2,\ell}|S^n,\W_1,\W_2,\W_3,\breve{\Y}_2^{\ell-1})$. Thus,
\begin{align}
&n(\tau R_{32}-\epsilon_{2n})
\leq \tau \sum_{\ell=1}^n I(W_{32};\Y_{2,\ell}|S^n,\W_1,\W_2,W_{31},\breve{\Y}_2^{\ell-1})\nonumber\\
&\quad= \tau \sum_{\ell=1}^n I(W_{32};\Y_{2,\ell}|S^n,\W_1,\W_2,W_{31},\breve{\Y}_2^{\ell-1},S_\ell=1)\nonumber,
\end{align}
since for a given $\ell$, this mutual information is equal to $I(W_{32};\H_{32}\X_{3,\ell}+\Z_{2,\ell}|S^n,\W_1,\W_2,W_{31},\breve{\Y}_2^{\ell-1})$ independent of the state $S_\ell$. Combining the two bounds yields
\begin{align}
&n(R_{31}+\tau R_{32}-\epsilon_{3n}-\tau o(\log(\rho)))\nonumber\\
&\leq \tau\sum_{\ell=1}^n I(\W_3;\Y_{2,\ell}|S^n,\W_1,\W_2,\breve{\Y}_2^{\ell-1},S_\ell=1)\\
&\leq n\tau M_3\log(\rho)+n\tau o(\log(\rho))
\end{align}
which follows using similar steps as in the proof of Lemma \ref{Lemma:DoF_Bounds}.
This leads to the desired result.
\end{IEEEproof}





Despite its simplicity, Lemma \ref{Lemma:Out} suffices for our purpose. Based on this lemma, the following theorem proves the necessity of adaptation in the intermittent case.

\begin{theorem}
For an intermittent 3WC with $M_1>M_3$, $\mathcal{D}_{\rm a,\tau}\not\subset\mathcal{D}_{\rm r,\tau}^{[\rm out]}$, and hence adaptation is necessary.
\end{theorem}
\begin{IEEEproof}
It suffices to show that $\exists\d\notin\mathcal{D}_{\rm r,\tau}^{[\rm out]}$ which is achievable using an adaptive scheme. To is end, suppose that only node 3 has a message to node 1, while node 2 acts as a relay to support node 3 which employs block-Markov encoding. Consider $B$ transmission blocks, each consisting of $n$ channel uses, and let $a_2,a_3\in[0,1]$, be chosen so that $a_2M_2,a_3M_3\in\mathbb{N}$. In block 1, node 3 encodes a message $w_{31}(1)$ to a codeword $\x_{31}^n$ with $\x_{31,\ell}\in\mathbb{C}^{a_3M_3}$, and sends it to node 1 using $a_3M_3$ antennas. Node 1 receives only $m$ symbols corresponding to $s_{1,\ell}=1$ where $s_1^n$ is the state sequence in this block, with $m\leq n$ and $\frac{m}{n}\approx\tau$ as $n$ grows. However, node 2 receives all symbols, and thus, obtains $n-m$ codeword symbols from node 3 that have not been received by node 1. In block 2, node 3 sends $w_{31}(2)$ similar to block 1, while nodes 2 cooperates with node 3. It does so by multiplying the received signal in block 1 by $\H_{32}^\dagger$ to obtain a noisy version of $\x_{31}^n$ given by $\vu_{31}^n=\x_{31}^n+\tilde{\z}_2^n$ where $\tilde{\z}_{2,\ell}$ consists of $a_3M_3$ components of $\H_{32}^\dagger\z_{2,\ell}$, and then constructing $\v_{31}^m$ out of $\vu_{31,\ell}$ with $\ell\in\{t\in\{1,\ldots,n\}|s_{1,t}=0\}$, where $\v_{31,\ell}\in\mathbb{C}^{a_2M_2}$. Then, it sends a new symbol of $\v_{31}^m$ to node 1 in transmission $\ell$ if $s_{2,\ell-1}=1$, and repeats the previously transmitted symbol otherwise. This construction requires $ma_2M_2\leq (n-m)a_3M_3$. The signal $\v_{31}^m$ is sent from node 2 so that it is received linearly independent of $\x_{31}^n$ at node 1. Thus, node 1 receives a total of $m a_2M_2+m a_3M_3$ symbols in this block if $a_2M_2+a_3M_3\leq M_1$. At the end of this block, node 1 is able to decode $w_{31}(1)$ by combining its received signals from blocks 1 and 2. The same is repeated over blocks $3,\ldots,B-1$. In block $B$, only node 2 is active and delivers $m a_2M_2$ symbols to node 1. The achievable DoF is the ratio of the total number of delivered symbols to the total number of transmissions, i.e.,
\begin{align*}
d_{31}=\frac{(B-1)(m a_3M_3+m a_2M_2)}{nB}\approx\tau a_3M_3+\tau a_2M_2,
\end{align*}
for large $n$ and $B$. The constraints from above are
\begin{align}
0\leq a_2,a_3&\leq 1,&a_2M_2,a_3M_3&\in\mathbb{N},\\
a_2M_2+a_3M_3&\leq M_1,&\tau a_2M_2-\bar{\tau}a_3M_3&\leq0.
\end{align}
Now, we need to maximize $d_{31}$ with respect to $a_2$ and $a_3$ subject to these constraints. A feasible solution can be obtained as follows. First, we ignore the second constraint, which leads to a linear program which can be solved using the simplex method \cite{MatousekGartner}. Solving the linear program leads to $a_3^*=1$ and $a_2^*=\min\left\{\frac{M_1-M_3}{M_2},1,\frac{\bar{\tau}M_3}{\tau M_2}\right\}$. Then, we round $a_2^*M_2$ and $a_3^*M_3$ down to the nearest integer to obtain $a_3=1$, and $a_2M_2=\min\left\{M_1-M_3,M_2,\left\lfloor\frac{\bar{\tau}M_3}{\tau}\right\rfloor\right\}$. This leads to the achievability of $\min\left\{\tau M_1,\tau M_2+\tau M_3, \tau M_3+\tau\left\lfloor\frac{\bar{\tau}M_3}{\tau}\right\rfloor\right\}\triangleq d_{31,\rm a}$. Thus, the DoF tuple $\d_{\rm a}=(0,0,0,0,d_{31,\rm a},0)\notin\mathcal{D}_{\rm r,\tau}^{[\rm out]}$ is achievable, which proves the desired result.
\end{IEEEproof}

This theorem proves the {\it necessity of adaptation in the intermittent case}, where cooperation between nodes 2 and 3 becomes necessary to achieve higher DoF.

\section{Conclusion}
\label{Sec:Conclusion}
In this paper, we have investigated the impact of intermittency on the DoF region of the MIMO three-way channel. We have seen that adaptive encoding, can be either necessary or not, depending on the performance criterion. As far as the sum-DoF is concerned, adaptive encoding is not necessary, and the optimal sum-DoF can be achieved with restricted (nonadaptive) encoding. Since the sum-DoF might be unfair in terms of per-user or per-stream DoF, the DoF region is of high importance. In this case, we have shown that adaptation is in fact necessary, and that collaboration between nodes using adaptive encoding enlarges the DoF region beyond what can be achieved using adaptive encoding. 

\end{document}